\documentclass[journal]{IEEEtran}

\pdfoutput=1

\usepackage{amsmath,bm,amssymb,amsfonts,amsthm}
\usepackage[linesnumbered, ruled, vlined]{algorithm2e}
\usepackage{graphicx}
\usepackage{xcolor} 
\usepackage{relsize}
\usepackage{textcomp}
\usepackage{listings}

\usepackage{changepage}
\usepackage{hhline}
\usepackage{multirow}
\usepackage{nomencl}
\usepackage{mathtools}
\usepackage{commath}
\usepackage{booktabs}
\usepackage{subfiles}
\usepackage{tabularx}
\usepackage{lscape}
\usepackage{rotating}

\usepackage[normalem]{ulem}
\usepackage[utf8]{inputenc}
\usepackage[hyphens]{url}
\usepackage{lineno}
    \modulolinenumbers[5]
\usepackage[caption=false]{subfig}
\usepackage{enumitem}
    \setlist{nolistsep}
\usepackage{gensymb}
\usepackage[hidelinks]{hyperref}
\usepackage{makecell}

\usepackage{scalerel}
\usepackage{tikz}
\usetikzlibrary{svg.path}
\definecolor{orcidlogocol}{HTML}{A6CE39}
\tikzset{
  orcidlogo/.pic={
    \fill[orcidlogocol] svg{M256,128c0,70.7-57.3,128-128,128C57.3,256,0,198.7,0,128C0,57.3,57.3,0,128,0C198.7,0,256,57.3,256,128z};
    \fill[white] svg{M86.3,186.2H70.9V79.1h15.4v48.4V186.2z}
                 svg{M108.9,79.1h41.6c39.6,0,57,28.3,57,53.6c0,27.5-21.5,53.6-56.8,53.6h-41.8V79.1z M124.3,172.4h24.5c34.9,0,42.9-26.5,42.9-39.7c0-21.5-13.7-39.7-43.7-39.7h-23.7V172.4z}
                 svg{M88.7,56.8c0,5.5-4.5,10.1-10.1,10.1c-5.6,0-10.1-4.6-10.1-10.1c0-5.6,4.5-10.1,10.1-10.1C84.2,46.7,88.7,51.3,88.7,56.8z};
  }
}
\newcommand\orcidicon[1]{\href{https://orcid.org/#1}{\mbox{\scalerel*{
\begin{tikzpicture}[yscale=-1,transform shape]
\pic{orcidlogo};
\end{tikzpicture}
}{|}}}}
\usepackage[noadjust]{cite}
\begin{document}

\title{\huge Fault Current-Constrained Optimal Power Flow \\ on Unbalanced Distribution Networks}

\author{
    Jose~E.~Tabarez $^{1}$\orcidicon{0000-0003-4800-6340},
    Arthur~K.~Barnes $^{1}$\orcidicon{0000-0001-9718-3197},
    Adam~Mate $^{1}$\orcidicon{0000-0002-5628-6509}, and
    Russell~W.~Bent $^{1}$\orcidicon{0000-0002-7300-151X}
    \vspace{-0.1in}

\thanks{Manuscript submitted:~May.~31,~2022. 
Current version:~Aug.~20,~2022.
}

\thanks{$^{1}$ The authors are with the Advanced Network Science Initiative at Los Alamos National Laboratory, Los Alamos, NM 87545 USA. Email: jtabarez@lanl.gov, abarnes@lanl.gov, amate@lanl.gov, rbent@lanl.gov.}

\thanks{LA-UR-22-22638. Approved for public release; distribution is unlimited.}

}

\markboth{2022 IEEE ISGT ASIA -- 11th International Conference on Innovative Smart Grid Technologies, November~2022}{}

\maketitle

% ==================================%
%     P A P E R   C O N T E N T     % 
% ==================================%

\begin{abstract}
With the proliferation of distributed generation into distribution networks, the need to consider fault currents in the dispatch problem becomes increasingly relevant.
This paper introduces a method for adding fault current constraints into optimal power flow in order to reduce fault currents while minimizing generation cost.
The optimal power flow problem is formulated as a single optimization problem with sub-networks representing the faults of interest. Having a single optimization problem allows the decision variables to be coupled across the optimal power flow and the fault current studies without having to iterate over possible solutions.
The proposed method is applicable to unbalanced distribution networks, including those with transformers that introduce phase-shifts.
\end{abstract}


%%%%%%%%%%%%%%%%%%%%%%%%%%%%%%%%%%%%%%%%%%
\section{Introduction} \label{sec:introduction}
\indent

The proliferation of distributed generation (DG) into distribution networks improves resilience of the entire electrical grid, but these additions also present challenges for protection. Networks with DG experience highly varying fault current magnitudes and directions based on the placement, type of technology, and operation of the generation from the original design of the distribution network.
The highly varying fault current magnitudes and directions are further exacerbated in the case of microgrids, where the fault current magnitudes and directions differ greatly between both grid-connected and islanded operating modes because of DG and multiple switching configurations. Therefore, it is reasonable to assert that faults current constraints should be added to operational decisions of the distribution network beyond just the optimal placement of distribution generation.
This paper introduces a method for fault current constrained optimal power flow (OPF) on unbalanced distribution networks, which is suitable for the online operation of such networks and can be extended to islanded microgrids as well.

\vspace{0.25in}
%%%%%%%%%%%%%%%%%%%%%%%%%%%%%%%%%%%%%%%%%%
\section{Background and Motivation}
\indent

DG can provide resilience to the distribution network through its many benefits \cite{PEPERMANS2005787, akorede2010review}, and research has been conducted to determine how best to exploit these benefits.
References \cite{akorede2010review, abookazemi2010review, georgilakis2013optimal} provide an overview of the state-of-the-art work that addresses the optimal placement of DGs in distribution networks in order to improve the voltage profile of feeders, decrease losses, increase security to critical loads, and provide other benefits (such as technical, economic, and environmental). Although \cite{martinez2009impact, meskin2020impact, alam2018fault} discuss some of the drawbacks of the optimal placement of DGs, adverse effects of fault currents in the distribution network are not fully addressed.
The purpose of this work is present a method of including fault currents in the OPF to determine the dispatch of DG subject to fault current constraints. Even though the inclusion of fault current constraints in OPF is not a novel concept \cite{vovos2005direct, vovos2005optimal, kim2021optimal, amirhossein2014optimal}, existing formulations rely on iterative methods to determine and include the fault currents into the OPF problem. The method presented in this paper, however, allows for the fault currents calculations to be directly incorporated into the OPF problem and applies constraints to them.

\vspace{0.25in}
%%%%%%%%%%%%%%%%%%%%%%%%%%%%%%%%%%%%%%%%%%
\section{Proposed Method}
\indent

The method proposed in this paper is a new formulation that is being added to PowerModelsProtection.jl (PMsP) \cite{barnes21-pmsp}, an optimization-based framework for short-circuit analysis.
This method considers short-circuit constrained OPF as a multi-network formulation: the first network represents the operation of the distribution network or microgrid under nominal pre-fault conditions, while additional networks represent the short-circuit behavior of the network under a collection of fault scenarios. Each additional network corresponds to a fault scenario, where a fault is described by the faulted node(s) and phasing(s) of the fault (e.g. line-ground, line-line, etc.).
While PMsP is sufficiently general to consider a scenario corresponding to multiple simultaneous faults, typically a scenario includes only a single fault.

PMsP is an addition to the family of open-source packages under the PowerModels.jl (PMs) \cite{coffrin-PMs} umbrella; PMs provide a platform to formulate and solve a variety of power systems optimization problems
Currently, PMsP only formulates mixed linear/nonlinear non-convex short-circuit formulations, and it has an additional dependency on PowerModelsDistribution.jl (PMsD) \cite{fobes-PMsD} for deriving the mathematical models and optimization based formulation for transmission and distribution networks, respectively.

Since the focus of this paper is on short-circuit faults in unbalanced distribution networks, the formulation and models will be based on the framework of PMsP, with mathematical modifications to the models and formulation to accommodate the short-circuit calculations.
The reason for the modifications is the fact that PMsD models the network and devices in the phase domain (three-wire Kron-reduction or neutral wire included), but the fault models of some devices are defined in symmetrical components. Additionally, settings for protective relays commonly rely on values based in symmetrical components.
PMsD provides the ability to formulate an optimization problem in either polar and rectangular form. PMsP utilizes the rectangular form to directly incorporate current variables, both real and imaginary, into the optimization problem.

\vspace{0.1in}
The following symbols are used in the below formulations:
\(\mathcal{B}\) is a set of buses in a network;
\(\mathcal{G}_{ref_b}\) is a set of reference voltage sources at bus $b$;
\(\mathcal{G}_b\) is a set of generators at bus $b$;
\(\Phi_b \subseteq \{A, B, C\}\) is the set of phases at bus $b$;
\(\mathcal{S}_b\) is a set of shunts at bus $b$;
\(\mathcal{T}_b\) is a set of transformers connected to bus $b$;
\(\mathcal{E}_b\) is a set of lines connected to bus $b$;
\(\mathcal{F}\) is a set of faults in network; and 
\(\mathcal{N}\) is a set of networks.

\subsection{Generator Models}
\indent

\normalsize For the case study, DG is modeled as constant-power generator in the OPF formulation and as voltage source behind an impedance for the short-circuit formulation. The original names of the generators -- i.e., PV1, Battery, Wind, etc. -- were kept.
Inverter models are included in PMsP, but were not used because of their low fault current injection \cite{kou2020fault}.
To highlight the fault current constraints, all inverters were converted to synchronous generation models,in which the maximum fault current contribution was set to be $K_f = 5$ times the rated current of the generator \cite{meskin2020impact}.

\vspace{0.1in}
\noindent
\normalsize 1) \textit{Constant-Power Generator OPF}:
\vspace{0.05in}

The constant-power generator model used in the OPF formulation is presented in Equations~(\ref{const:constant_gen}a)-(\ref{const:constant_gen}e).
When the bus voltage is above or below the bounds, the model reverts to a constant impedance model based on the dispatched power and voltage bound that was violated.
\vspace{0.05in}

\small
\noindent $\forall g \in {\cal G},~\forall b \in {\cal B},~\forall \phi \in \Phi_{b}$
\begin{subequations}\label{const:constant_gen}
\begin{align}
V^\phi_{r_g} \cdot I^\phi_{r_g} + V^\phi_{i_g} \cdot I^\phi_{i_g} &= P^\phi_{g} \\
V^\phi_{i_g} \cdot I^\phi_{r_g} - V^\phi_{r_g} \cdot I^\phi_{i_g} &= Q^\phi_{g}\\
\Delta P^\phi_{g} &\le 5\% \\
\Delta Q^\phi_{g} &\le 5\% \\
-\frac{P^\phi_{g}}{pf} \cdot \sin(\textrm{acos}(pf)) \le Q^\phi_{g} &\le \frac{P^\phi_{g}}{pf} \cdot \sin(\textrm{acos}(pf))
\end{align}
\end{subequations}

\vspace{0.05in}
\normalsize Constraints~(\ref{const:constant_gen}a) and (\ref{const:constant_gen}b) define the output power based on the terminal voltages and currents. Real power is a decision variable in the objective functions to minimize cost. The last set of constraints are operational constraints to ensure that the generator operates within the desired power factor (pf) range and that the power injected on each phase is balanced to within 5\% of the other phases.

\vspace{0.1in}
\noindent
\normalsize 2) \textit{Constant-Power Generator Short-Circuit}:
\vspace{0.05in}

For the short-circuit formulation, the constant-power generators are modeled as a voltage source behind an impedance. Both the voltage and impedance are determined from the operational state of the generator produced from the OPF.   
\vspace{0.05in}

\small
\noindent $\forall g \in {\cal G},~\forall b \in {\cal B},~\forall \phi \in \Phi_{b}$:
\begin{subequations}\label{const:gen_scConst}
\begin{align}
(V^\phi_{r_{opf}})^2 &= K_f \cdot (z^\phi_r \cdot P^\phi_{g} + z^\phi_i \cdot Q^\phi_{g}) \\
(V^\phi_{i_{opf}})^2 &= K_f \cdot (-z^\phi_r \cdot Q^\phi_{g} + z^\phi_i \cdot P^\phi_{g}) \\
V^\phi_{r} &= V^\phi_{r_{opf}} - z^\phi_r \cdot I^\phi_{r} + z^\phi_i \cdot I^\phi_{i} \\
V^\phi_{i} &= V^\phi_{i_{opf}} - z^\phi_r \cdot I^\phi_{i} - z^\phi_i \cdot I^\phi_{r}  
\end{align}
\end{subequations}

\vspace{0.05in} \normalsize
The variables \(P^\phi_{g}\), \(Q^\phi_{g}\), \(V^\phi_{r_{opf}}\) and \(V^\phi_{i_{opf}}\) are OPF decision variables that are shared with the short-circuit formulation and are used to determine the maximum fault current from the generator (current is based on the operating  point of the generator and scaled by $K_f$ \cite{meskin2020impact}). Equations (\ref{const:gen_scConst}c) and (\ref{const:gen_scConst}d) determine the fault current supplied by the generator as function of the operating voltage and series impedance.

\subsection{OPF Formulation}
\indent

\normalsize The optimization based OPF formulation implemented in PMsP is as follows:

\vspace{0.1in}
\noindent
\normalsize Objective Function: \\
\vspace{0.05in}
\small
\begin{equation}
Min~ \sum_{g \in G} \sum_{\phi \in \Phi} \left( C \cdot P^\phi_g \right)
\end{equation}
\vspace{0.05in}

\noindent
\normalsize Reference Voltage Sources Constraints: \\
\vspace{0.05in}
\small
\noindent $ \forall g \in \mathcal{G}_{ref_n}, ~\forall b \in {\cal B},~\forall \phi \in \Phi_{b}$
\begin{subequations}
\label{const:ref_volt_scr}
\begin{align}
&V^\phi_{r_g} = V^\phi_{g_{setp.}} \cdot cos \left (\theta_{g_{setp.}} \right )& \\
&V^\phi_{i_g} = V^\phi_{g_{setp.}} \cdot sin \left (\theta_{g_{set.}} \right )& \\
&V^\phi_{r_b} = V^\phi_{r_g} - r^\phi_g \cdot I^\phi_{r_g} + x^\phi_g \cdot I^\phi_{i_g}& \\
&V^\phi_{i_b} = V^\phi_{i_g} - r^\phi_g \cdot I^\phi_{i_g} - x^\phi_g \cdot I^\phi_{r_g}& 
\end{align}
\end{subequations}
\vspace{0.05in}

\noindent
\normalsize Generator Power Constraints: \\
\vspace{0.05in}
\small
\noindent $ \forall g \in \mathcal{G}_b, ~\forall b \in {\cal B},~\forall \phi \in \Phi_{b}$
\begin{align}
&Constraints~(\ref{const:constant_gen}a)-(\ref{const:constant_gen}e)&
\end{align}
\vspace{0.05in}

\noindent
\normalsize Kirchhoff's Current Constraints for Buses: \\
\vspace{0.05in}
\small
\noindent $\forall b \in {\cal B},~\forall \phi \in \Phi_{b}$
\begin{subequations}
\label{const:kirch_i_nonf}
\begin{align}
&\sum_{(b,j) \in \mathcal{E}_b} I^\phi_{r_{(b,j)}} + \sum_{(b,j) \in \mathcal{T}_b} I^\phi_{r_{(b,j)}}= \sum_{g \in \mathcal{G}_b} I^\phi_{r_{g}} + \sum_{s \in \mathcal{S}_b} I^\phi_{r_{s}} &\\
&\sum_{(b,j) \in \mathcal{E}_b}I^\phi_{i_{(b,j)}} + \sum_{(b,j) \in \mathcal{T}_b} I^\phi_{i_{(b,j)}}= \sum_{g \in \mathcal{G}_b} I^\phi_{i_{g}} + \sum_{s \in \mathcal{S}_b} I^\phi_{i_{s}} & 
\end{align}
\end{subequations}
\vspace{0.05in}

\noindent
\normalsize Voltage Drop Constraints: \\
\vspace{0.05in}
\small
\noindent $\forall (b,j) \in {\cal E},~\forall b \in {\cal B},~\forall \phi \in \Phi_{b}$
\begin{subequations}
\label{const:volt_drop}
\begin{align}
&V^\phi_{r_{b}} = V^\phi_{r_{b}} - r^\phi_{(b,j)} \cdot I^\phi_{r_{(b,j)}} + x^\phi_{(b,j)} \cdot I^\phi_{i_{(b,j)}}& \\
&V^\phi_{i_{b}} = V^\phi_{i_{b}} - r^\phi_{(b,j)} \cdot I^\phi_{i_{(b,j)}} - x^\phi_{(b,j)} \cdot I^\phi_{r_{(b,j)}}& \\
&V^\phi_{b_{min}} \le (V^\phi_{r_b})^2 + (V^\phi_{i_b})^2 \le V^\phi_{b_{max}} & \\
& (I^\phi_{r_{(b,j)}})^2 + (I^\phi_{i_{(b,j)}})^2 \le I^\phi_{(b,j)_{thermal}} &
\end{align}
\end{subequations}
\vspace{0.05in}

\newpage

\noindent
\normalsize Transformer Constraints: \\
\vspace{0.05in}
\small
\noindent $\forall (b,j) \in {\cal T},~\forall b \in {\cal B},~\forall \phi \in \Phi_{b}$
\begin{subequations}
\label{const:ofp_transformer_thermal}
\begin{align}
&W_{(b,j)} \cdot V^\phi_{r_{b}} = \eta_{(b,j)} \cdot V^\phi_{r_{j}} & \\
&W_{(b,j)} \cdot V^\phi_{i_{b}} = \eta_{(b,j)} \cdot V^\phi_{i_{j}} & \\
&I^\phi_{i_{b}} = \eta_{(b,j)} I^\phi_{i_{j}} & \\
& (I^\phi_{r_{(b,j)}})^2 + (I^\phi_{i_{(b,j)}})^2 \le I^\phi_{(b,j)_{thermal}} &
\end{align}
\end{subequations}
% \vspace{0.05in}

\vspace{0.1in} \normalsize
In general, the OPF formulation is similar to the short-circuit formulation described below, but with the following differences: fault constraint removed; constraints defining voltage/power/thermal limits included; and active/reactive powers defined as variables instead of set-points.

\subsection{Short-Circuit Formulation}

\normalsize The optimization based short-circuit current calculation formulation implemented in PMsP is as follows:
\vspace{0.1in}

\noindent
\normalsize Reference Voltage Sources Constraints: \\
\vspace{0.05in}
\small
\noindent $ \forall g \in \mathcal{G}_{ref,b}, ~\forall b \in {\cal B},~\forall \phi \in \Phi_{b}$
\begin{align}
\label{const:opf_ref_volt_scr}
&Constraints~(\ref{const:ref_volt_scr}a)-(\ref{const:ref_volt_scr}d)&
\end{align}
\vspace{0.05in}

\noindent
\normalsize Generator Power Constraints:\\
\vspace{0.05in}
\small
\noindent $ \forall g \in \mathcal{G}_b, ~\forall b \in {\cal B},~\forall \phi \in \Phi_{b}$
\begin{align}
&Constraints~(\ref{const:gen_scConst}a)-(\ref{const:gen_scConst}d)&
\end{align}
\vspace{0.05in}

\noindent
\normalsize Fault Current Constraints: \\
\vspace{0.05in}
\small
\noindent $ \forall f_b \in \mathcal{F}, ~\forall b \in {\cal B},~\forall \phi \in \Phi_{b}$
\begin{subequations}
\label{const:fault_current}
\begin{align}
I^\phi_{r_{f_b}} = \sum_{c \in \Phi} \left( G_{\phi,c} \cdot V^c_{r_b} \right) \\
I^\phi_{i_{f_b}} = \sum_{c \in \Phi} \left( G_{\phi,c} \cdot V^c_{i_b} \right)
\end{align}
\end{subequations}
\vspace{0.05in}

\noindent
\normalsize Kirchhoff's Current Constraints for Unfaulted Buses: \\
\vspace{0.05in}
\small
\noindent $\forall b \in {\cal B},~\forall \phi \in \Phi_{b}$
\begin{align}
&Constraints~(\ref{const:kirch_i_nonf}a)-(\ref{const:kirch_i_nonf}b)&
\end{align}
\vspace{0.05in}

\noindent
\normalsize Kirchhoff's Current Constraints for Faulted Buses: \\
\vspace{0.05in}
\noindent $\forall b \in {\cal B},~\forall \phi \in \Phi_{b}$
\begin{subequations}\label{const:kirch_i_f}
\begin{align}
&\sum_{b,j \in \mathcal{E}_b} I^\phi_{r_{(b,j)}} + \sum_{b,j \in \mathcal{T}_b} I^\phi_{r_{(b,j)}}= \sum_{g \in \mathcal{G}_b} I^\phi_{r_{g}} + \sum_{s \in \mathcal{S}_b} I^\phi_{r_{s}} + I^\phi_{r_{f}}&\\
&\sum_{b,j \in \mathcal{E}_b} I^\phi_{i_{(b,j)}} + \sum_{b,j \in \mathcal{T}_b} I^\phi_{i_{(b,j)}}= \sum_{g \in \mathcal{G}_b} I^\phi_{i_{g}} + \sum_{s \in \mathcal{S}_b} I^\phi_{i_{s}} + I^\phi_{i_{f}}& 
\end{align}
\end{subequations}
\vspace{0.05in}

\noindent
\normalsize Voltage Drop Constraints: \\
\vspace{0.05in}
\small
\noindent $\forall (b,j) \in {\cal E},~\forall b \in {\cal B},~\forall \phi \in \Phi_{b}$
\begin{align}
&Constraints~(\ref{const:volt_drop}a)-(\ref{const:volt_drop}b)& 
\end{align}
\vspace{0.05in}

\noindent
\normalsize Transformer Constraints: \\
\vspace{0.05in}
\small
\noindent $\forall (b,j) \in {\cal T},~\forall b \in {\cal B},~\forall \phi \in \Phi_{b}$
\begin{align}
\label{const_transformer_sc}
&Constraints~(\ref{const:ofp_transformer_thermal}a)-(\ref{const:ofp_transformer_thermal}d)&
\end{align}
%\vspace{0.05in}

\vspace{0.1in} \normalsize
The general formulation does not require an objective function as the fault currents are represented by Constraints~(\ref{const:fault_current}a)-(\ref{const:fault_current}b) within the problem with the use of a fault admittance matrix, which represents the fault impedances between phases and ground.
Faults involving impedances between phases and ground cannot be directly modeled. For these types of faults a star-mesh transformation is used to define the equivalence impedance as an admittance matrix \cite{barnes21-pmsp}.

\vspace{0.05in}
Constraints~(\ref{const:ref_volt_scr}a)-(\ref{const:ref_volt_scr}d) define the reference voltage sources in the network.
In general, the distribution substation would be the main voltage source when the network is operating in grid-connected mode. During a fault, the substation is modeled as a voltage source behind an impedance; this impedance is derived from the three-phase and single-phase short-circuit powers (\ref{const:ref_volt_scr}b).
However, the model used for substations is not adequate for representing the behavior of inverter-interfaced generation in microgrids. PMsP provides a voltage-power generator model that is designed to represent the short-circuit behavior of inverters more accurately \cite{barnes21-pmsp, barnes22-dse-ldbus}.

\subsection{Fault Current-Constrained OPF}
\vspace{-0.1in}
\indent

\normalsize
PMsP is capable of creating a network of optimization problems and deriving coupling constraints to link the individual optimization problems into a single optimization problem.
For this paper, the protection-constrained OPF is formulated as a problem that minimizes the costs of generation in the OPF network ($ n = 0$) subject to fault current limits in the short-ciruit networks ($n \in \mathcal{N}, \; n \ne 0$). 
The first network is the OPF problem and all other networks are the short-circuit problems. The formulation is as follows:   

\vspace{0.1in}
\noindent
\normalsize Objective Function: \\
\vspace{0.05in}
\small
\begin{align}\label{obj:opf_fs}
&Min~ \sum_{g \in G_0} \sum_{\phi \in \Phi} \left( C \cdot P^\phi_g \right)  + \sum_{f \in F} \sum_{\phi \in \Phi}I^\phi_{f_b}&
\end{align}
\vspace{0.05in}

\noindent
\normalsize OPF Constraints: \\
\vspace{0.05in}
\small
\noindent $ for~n = 0 $:
\begin{align}
&Constraints~(\ref{const:constant_gen}a)-(\ref{const:constant_gen}e)& \\
&Constraints~(\ref{const:ref_volt_scr})-(\ref{const:ofp_transformer_thermal})& 
\end{align}
\vspace{0.05in}

\noindent
\normalsize Faulted Network Constraints: \\
\vspace{0.05in}
\small
\noindent $\forall n \in {\cal N},n\ne 0$:
\begin{align}
&Constraints~(\ref{const:gen_scConst}a)-(\ref{const:gen_scConst}d)& \\
&Constraints~(\ref{const:opf_ref_volt_scr})-(\ref{const_transformer_sc})& \\
&I_{f_{n_b}} = (I^\phi_{r_{f_b}})^2 + (I^\phi_{i_{f_b}})^2& 
\end{align}
%\vspace{0.05in}

\vspace{0.05in} \normalsize
In the case study, the cost of power and the fault currents are scaled in order to be included into one objective function. With the scaling, both the cost and the fault current values in the objective function range from 0 to 2; this eliminates the creation of a dual optimization problem.

\section{Case Study System}
\indent

\begin{figure*}[!htbp]
\centering
\includegraphics[width=0.7\textwidth]{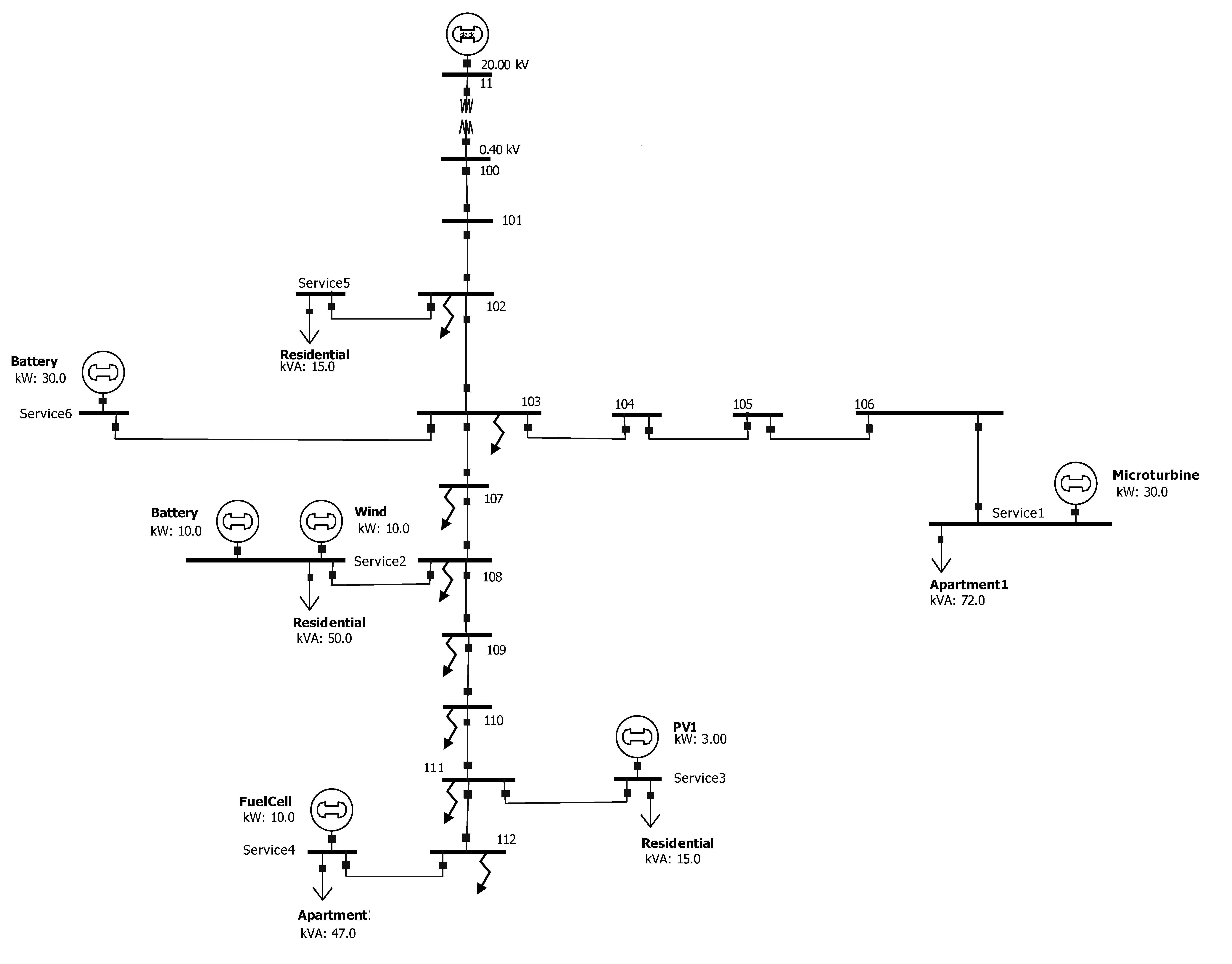}
\caption{Cigre LV Network \cite{papathanassiou2005benchmark}. Fault locations are illustrated with jagged arrows.}
\label{fig:cigre_network}
\end{figure*}

The case study system employed here is a modified version of the CIGRE low-voltage microgrid case study system \cite{papathanassiou2005benchmark}, illustrated in Fig.~\ref{fig:cigre_network}. It was designed to be representative of a low-voltage (400~V line-line) distribution feeder -- typical of Asian, European, or African secondary residential distribution networks -- and to be used for microgrid protection case studies; it includes both DG and grounding resistances.

Key parameters of the system are as follows: total load power ($S_d$) is 199~kVA; total generator power ($S_g$) is 93~kVA; grid frequency ($f$) is 50~Hz; number of lines is 18; number of loads is 5; number of generators is 7; load power factor ($p_f$) is 0.90; the maximum distance from the substation ($d$) is 385~meters.
As stated above, all DGs were modeled as synchronous generators and their costs were arbitrarily set to \$0.80~kW/h, and a 24-hour operating period is considered.

\subsection{Procedures to Validate Method} \label{sec:validation}
\vspace{-0.1in}
\indent

To validate the proposed method and provide insight into the solution, several pre-optimization analyses were performed.
The first analysis was power flow (PF), which provided the maximum power supplied by the substation and bus voltages to provide a reference for the OPF solutions.
The next analysis was a basic OPF (min cost) to determine the set points of the DGs, which provides a reference to the optimal cost without the consideration of fault currents.
The final analyses were fault studies on the test case with and without DGs; in the case of DG, the set points of the DG were based on the results from the basic OPF. Theses fault studies provide the minimum and maximum faults currents that could exist in the case study.

\subsection{Fault Current Constrained OPF} \label{sec:protection-constrained-opf}
\vspace{-0.1in}
\indent

To demonstrate the implemented protection-constraint OPF formulation, a problem was created to minimize the cost of generation while trying to reduce fault currents in the network.
Based on the structure of the problem, the only way the optimization problem can reduce the fault currents is by decreasing the setpoints of the DG in the model.
For this case study, fault scenarios consisting of three-phase grounded faults at Buses 102, 103, and 107 -- 112 are considered, but the formulation does not limit the number or type of faults.

%%%%%%%%%%%%%%%%%%%%%%%%%%%%%%%%%%%%%%%%%%
\section{Results}
\indent

Table~\ref{tab:opendss-fault-current-comparison} presents the bus voltage results of the case study for the different analyses and the fault current constraint OPF formulation.
It can be seen that the case study requires DG in order to maintain bus voltages over \(0.9\) pu. The OPF solutions were constrained to maintain bus voltages above \(0.9\) pu, which both OPFs were able to satisfy with the addition of distributed generation.
In the fault current constrained OPF, the optimization took into account the fault current, which resulted in the OPF limiting the DG dispatch, causing the voltages at the buses to just satisfy the voltage magnitude limit constraint.

\begin{table}[!htbp]
\smaller
\centering
\caption{Bus Voltages A-Phase}
\begin{tabular}{|r|r|r|r|r|}
\hline
\textit{Bus} & \textit{PF (No-DG)} & \textit{OPF (DG)} & \textit{FC-OPF}  \\
\hline \hline
11  &   $1.000\; \angle0^{\circ}$ &   $1.000\; \angle0^{\circ}$ &   $1.000\; \angle0^{\circ}$   \\
100 &   $0.9091\; \angle-29.92^{\circ}$ &   $0.9879\; \angle-29.86^{\circ}$ &   $0.9829\; \angle-29.89^{\circ}$ \\
101 &   $0.9611\; \angle-29.79^{\circ}$ &   $0.9810\; \angle-29.75^{\circ}$ &   $0.9730\; \angle-29.76^{\circ}$ \\
102 &   $0.9470\; \angle-29.66^{\circ}$ &   $0.9742\; \angle-29.63^{\circ}$ &   $0.9631\; \angle-29.63^{\circ}$ \\
103 &   $0.9339\; \angle-29.54^{\circ}$ &   $0.9683\; \angle-29.51^{\circ}$ &   $0.9543\; \angle-29.50^{\circ}$ \\
104 &   $0.9253\; \angle-29.39^{\circ}$ &   $0.9639\; \angle-29.41^{\circ}$ &   $0.9495\; \angle-29.37^{\circ}$ \\
105 &   $0.9168\; \angle-29.26^{\circ}$ &   $0.9595\; \angle-29.30^{\circ}$ &   $0.9447\; \angle-29.25^{\circ}$ \\
106 &   $0.9083\; \angle-29.10^{\circ}$ &   $0.9550\; \angle-29.19^{\circ}$ &   $0.9399\; \angle-29.12^{\circ}$ \\
107 &   $0.9296\; \angle-29.46^{\circ}$ &   $0.9361\; \angle-29.44^{\circ}$ &   $0.9465\; \angle-29.42^{\circ}$ \\
108 &   $0.9180\; \angle-29.38^{\circ}$ &   $0.9579\; \angle-29.37^{\circ}$ &   $0.9387\; \angle-29.35^{\circ}$ \\
109 &   $0.9135\; \angle-29.34^{\circ}$ &   $0.9546\; \angle-29.33^{\circ}$ &   $0.9344\; \angle-29.30^{\circ}$ \\
110 &   $0.9091\; \angle-29.29^{\circ}$ &   $0.9513\; \angle-29.29^{\circ}$ &   $0.9300\; \angle-29.26^{\circ}$ \\
111 &   $0.9046\; \angle-29.24^{\circ}$ &   $0.9481\; \angle-29.22^{\circ}$ &   $0.9224\; \angle-29.18^{\circ}$ \\
Svc1    &   $0.8966\; \angle-28.82^{\circ}$  &  $0.9490\; \angle-29.00^{\circ}$  &   $0.9335\; \angle-28.92^{\circ}$ \\
Svc2    &   $0.9095\; \angle-29.18^{\circ}$  &  $0.9534\; \angle-29.23^{\circ}$  &   $0.9304\; \angle-29.15^{\circ}$ \\
Svc3    &   $0.8940\; \angle-28.94^{\circ}$  &  $0.9402\; \angle-29.01^{\circ}$  &   $0.9153\; \angle-28.93^{\circ}$ \\
Svc4    &   $0.8850\; \angle-28.88^{\circ}$  &  $0.9536\; \angle-28.96^{\circ}$  &   $0.9099\; \angle-28.86^{\circ}$ \\
Svc5    &   $0.9369\; \angle-29.39^{\circ}$  &  $0.9644\; \angle-29.36^{\circ}$  &   $0.9532\; \angle-29.36^{\circ}$ \\
Svc6    &   $0.9334\; \angle-29.54^{\circ}$  &  $0.9764\; \angle-29.63^{\circ}$  &   $0.9616\; \angle-29.59^{\circ}$ \\
\hline
\end{tabular}
\label{tab:opendss-fault-current-comparison}
\end{table}

\vspace{-0.05in}
Table~\ref{tab:opf_fs_fault_currents} illustrates the fault currents for the case study.
The "No-DG" case had the lowest fault current levels, due the substation being the only source. The maximum fault current occurred in the case where the DG was at maximum. These two cases provide boundaries for the fault current constrained OPF, and the table shows that the optimal point returned was indeed between the two boundaries. 

\begin{table}[!htbp]
\smaller
\centering
\caption{Fault Currents for Case Study System (Phase A Currents)}
\begin{tabular}{|r|r|r|r|}
\hline
\textit{Bus} & \textit{Power Flow (No-DG)} &  \textit{OPF (DG)} & \textit{FC-OPF} \\
\hline \hline
102 & 5967 A & 6639 A & 6424 A  \\
103 & 4708 A & 5403 A & 5184 A \\
107 & 3888 A & 4437 A & 4192 A  \\
108 & 3311 A & 3378 A & 3522 A  \\
109 & 2883 A & 3256 A & 3038 A  \\
110 & 2553 A & 2864 A & 2671 A  \\
111 & 2290 A & 2559 A & 2383 A  \\
112 & 2077 A & 2310 A & 2152 A  \\ \hline
Total & 27677 A & 30846 A & 29566 A  \\
\hline
\end{tabular}
\label{tab:opf_fs_fault_currents}
\end{table}

\vspace{-0.05in}
Table~\ref{tab:protection-constained-opf-fault-currents} provides the set points for distributed generation in the case study for each analysis and the fault current constrained OPF.

\begin{table}[!htbp]
\smaller
\centering
\caption{Distributed Generation for Case Study System (Phase A Currents)}
\begin{tabular}{|r|r|r|r|}
\hline
\textit{DG} & \textit{Power Flow (No-DG)} &  \textit{OPF (DG)} & \textit{FC-OPF} \\
\hline \hline
Microturbine &  0 kW & 30 kW & 30 kW  \\
Wind &  0 kW  & 10 kW & 0kW \\
Battery &  0 kW & 30 kW & 30 kW  \\
PV1 &  0 kW & 3 kW & 0 kW  \\
FuelCell &  0 kW & 10 kW & 0 kW  \\
PV2 &  0 kW & 10 kW & 0 kW  \\ \hline
Total Cost & \$0.00  & \$74.40  & \$48.00 \\
\hline
\end{tabular}
\label{tab:protection-constained-opf-fault-currents}
\end{table}

% \begin{table}[!htbp]
% \centering
% \caption{Distributed Generation Set-points for Case Study System (Phase A)}
% \begin{tabular}{|r|r|r|r|r|}
% \hline
% \textit{Gen} & \textit{P OPF} & \textit{Q OPF} &  \textit{P P-C OPF}&  \textit{Q P-C OPF}   \\
% \hline \hline
% Fuelcell & 10 kW & $-6.57$ kVAR & $0$ kW & $0.00$ kVAR  \\
% PV1 &  $3$ kW & $-2.19$ kVAR & $3$ kW & $-0.05$ kVAR\\
% $\mu$turbine &  $30$ kW & $9.34$ kVAR & $30$ kW & $9.37$ kVAR \\
% PV2 &  $10$ kW & $-6.57$ kVAR & $10$ kW & $-0.48$ kVAR \\
% Battery & $30$ kW & $-19.72$ kVAR & $30$ kW & $-19.72$ kVAR \\
% Wind & $10$ kW & $-6.75$ kVAR & $10$ kW & $-0.48$ kVAR \\
% \hline
% \end{tabular}
% \label{tab:protection-constrained-opf-gen-sp}
% \end{table}

%%%%%%%%%%%%%%%%%%%%%%%%%%%%%%%%%%%%%%%%%%
\section{Conclusions}
\indent

The concept of a short-circuit OPF has previously been demonstrated for balanced transmission and distribution networks; however, because of their iterative nature, these methods are limited in their ability to handle significant topological or operational changes to the network.
This work demonstrated the feasibility of applying the concept to unbalanced distribution networks and sub-networks with the full short-circuit current flow encoded as constraints. In doing so, the method is applicable to generalized unit commitment, where certain generator commitment can take into account multiple fault currents created by the OPF commitment problem.

In the case study, the fault current constrained OPF was able to limit the fault current by adjusting the DG set points to redistribute power while assuring load is supplied, as shown in Tables~\ref{tab:opendss-fault-current-comparison} - \ref{tab:protection-constained-opf-fault-currents}. 
The protection-constrained OPF optimization determined that the generations on buses ``Service1'' (a.k.a ``Svc1'') and ``Service6'' (a.k.a ``Svc6'') should be ON to reduce the cost of generation while limiting the fault current at the fault locations. 
This result is reasonable since the operational cost of the generation was relatively close as they utilized the same technology, but the electrical distance between the generation sources on buses `Svc1'' and ``Svc6'' and the fault location are further than the other generation sources resulting in less injected fault current.

Future work will continue to extend the presented formulation into microgrid protection, and develop relaxed convex approximations for the short-circuit models for protection equipment and grid-forming/grid-following inverters.

% ============================%
%     R E F E R E N C E S     % 
% ============================%

% \newpage
\bibliographystyle{unsrt}
\bibliography{references}

\end{document}